\documentclass[12pt,thmsa]{article}
\usepackage{sw20aip}

\input{tcilatex}
\begin{document}

\author{Zhenying Wen \\
Physics Department of Lanzhou University, Lanzhou 730000, China \and Hong
Zhao$^{*}$ \\
Physics Department of Xiamen University, Xiamen 361005, China}
\title{Mechanical Resonance of embedded clusters}
\date{The Date }
\maketitle

\begin{abstract}
Embedded clusters, which are embedded in bulk materials and different from
the surroundings in structures, should be common in materials. This paper
studies resonance of such clusters. This work is stimulated by a recent
experimental observation that some localized clusters behavior like fluid at
the mesoscopic scale in many solid materials [Science in China(Series B).
46, 176 (2003)]. We argue that the phenomenon is just a vivid illustration
of resonance of embedded clusters, driven by ubiquitous microwaves. Because
the underlying mechanism is fundamental and embedded structures are usual,
the phenomenon would have great significance in material physics.

PACS numbers:
\end{abstract}

Mechanical resonance is a common thread which runs through almost every
branch of physics. It occurs when the driving frequency matches the natural
frequency of the sample. Recently, mechanical resonance has been observed at
the mesoscopic scale for some nano-size materials\cite{1,2,3,4,5}. It takes
place in near-field scanning optical microscope probes\cite{1,2} and some
nanostructures such as carbon nanotubes\cite{3}, ZnO nanobelts\cite{4} and
SiO$_{\text{2}}$ nanowires\cite{5}. Mechanical resonance is a basic concept
in textbooks and it is known that the boundary condition determines the
normal modes of the samples. But theoretical analyses only cover samples
with unambiguous boundaries, i.e., two fixed ends, two free ends, or one
fixed end and one free end. Systematic knowledge is still lacking in such a
typical case: embedded clusters, as shown in Fig. 1(a), which are embedded
in a bulk material and are different from the surrounding parts in
structure, and so have boundaries neither fixed nor free. This kind of
clusters is usual at both macroscopic and mesoscopic scale in real world,
such as materials embedded in other materials and the metastable structures
in solids. It is clear that resonance can occur when the mechanical property
of the embedded cluster is extremely different from that of surrounding
materials and it can't occur when the properties of the embedded cluster and
the surrounding materials are quite similar. For general case, whether
resonance can occur and how it occurs are still unclear.

The study of this problem is motivated by a recent experimental discovery of
Gao $et$ $al.$\cite{6,7,8}. With an optical microscope, Gao $et$ $al.$ have
observed some localized clusters in the range of 0.1$\times $0.1$\mu $m$^2$
to 2$\times $2$\mu $m$^2$ on the surface of a sample of Cu-Zn-Al alloy under
normal pressure and temperature\cite{6,7}; these clusters behave like fluid:
amorphism, flowing, waving, and they also expand and then contract; such
motion can last from several seconds to several weeks\cite{6,7,8}. Later the
similar phenomena have been observed in some other solid alloys, mineral
crystal, gabbro, monocrystal nickel sulfate, semiconductor\cite{8}. In
short, they find mesoscopic-scale clusters in many materials which show
irregular movement at mesoscopic scale while their surroundings keep static.
This is an interesting phenomenon and we believe that after the underlying
mechanism is elucidated it will be recognized to be of significance and
intensive studies will be focused. Our viewpoint is that the phenomenon is
just a vivid illustration of resonance of the embedded clusters. The
fluid-like clusters are metastable structures in solids and just embedded
clusters referred above; ubiquitous microwaves can drive them into resonant
vibrations when microwave frequencies match their natural frequencies; the
resonant vibrations can become marked movements at the mesoscopic scale if
the sizes of the clusters and microwave intensities are appropriate.

For the sake of simplicity, we first employ a lattice as shown in Fig. 1(b)
to study the resonance of the embedded cluster in a 1D sample. The
Hamiltonian is 
\[
H=\sum_{i=1}^N\left[ \frac{p_i^2}2+\frac 12\kappa _i\left(
x_{i+1}-x_i\right) ^2+\frac 14\kappa _i\left( x_{i+1}-x_i\right) ^4\right] 
\]
where $p_i$ is the momentum of the $ith$ atom, and $x_i$ is the displacement
from the equilibrium position. The mass $m_i$ is set to unity and the first
and last atoms are fixed. The force constant $\kappa _i$ serves as the
control parameter, which takes the value $\kappa _o$ for $i\in [\frac{N-N_1}%
2+1,\frac{N+N_1}2]$ and is set to unity for the other atoms. Then the middle
part of the lattice with the $N_1$ atoms is structurally different from the
other parts of the lattice and thus mimics the embedded cluster. The other
parts represent the bulk sample. Notice that if $\kappa _i$ is a constant
for all atoms we will obtain the well-known Fermi-Pasta-Ulam model, which is
a paradigm model in lattice studies.

For the embedded cluster, it is obvious that resonance can occur in the
limit case of $\kappa _o\rightarrow 0$ (corresponding to fixed boundary
condition) or $\kappa _o\rightarrow \infty $ (to free boundary condition)
when the driving frequency matches its nature frequency. As we know, the
natural frequency is $\omega _n=\frac{n\pi }{N_1}\sqrt{\frac{\kappa _o}m}%
\cos \left( \frac{n\pi }{2N_1}\right) $ $\left( 1\right) $ for both limit
fixed and limit free boundary conditions, and the normal mode is $x_n=A\sin
(n\pi x_i/N_1)$ for the former and is $x_n=A\cos (n\pi x_i/N_1)$ for the
latter, where $x_i\in [\frac{N-N_1}2+1,\frac{N+N_1}2]$. On the other hand,
in the limit of $\kappa _o\rightarrow 1$ the embedded cluster appears
indistinguishable with the other parts.

We apply numerical calculations to test whether resonance can occur in the
embedded cluster in general case. In our simulations the first and last
fifty atoms are coupled to Langevin thermostats\cite{9}, which mimics the
bulk sample in the environment temperature. The temperature $T$ has been set
equal to $0.01$, corresponding to the room temperature\cite{10}. The
thermodynamic equilibrium state is established after sufficient long
evolutions, around $t\sim 10^5$. Then we apply periodic forces $F=f\cos
\left( \omega t\right) $ to the first fifty atoms, representing the driving
forces applied on the surface of the bulk sample. In fig. 2(a) we show the
average amplitude $A$ of the embedded cluster versus the driving frequency $%
\omega $ in the case of $N=3000$, $N_1=1000$ and $\kappa _o=0.1$. One can
see, there are two maximum points at $\omega _1=0.90\times 10^{-3}$ and $%
\omega _2=1.70\times 10^{-3}$. From equation $\left( 1\right) $ we know that
the first harmonic is $\omega _1^{^{\prime }}=1.0\times 10^{-3}$ and the
second harmonic is $\omega _2^{^{\prime }}=2.0\times 10^{-3}$ in theory for
the embedded cluster with limit fixed or free boundary condition. The two
maximum points $\omega _1$ and $\omega _2$ are close to the first and second
harmonics respectively. Fig. 2(b) displays the vibration pattern of the
lattice at $\omega _1$, and it is obvious that the pattern corresponds to
the first normal mode of the embedded cluster in limit fixed boundary
condition. We vary the value of $\kappa _o$ to $\kappa _o=2$. Resonance
occurs at $\omega =4.6\times 10^{-3}$, which is close to the theoretical
value $\omega ^{^{\prime \prime }}=4.4\times 10^{-3}$. The vibration pattern
(shown in Fig 2.(c)) agrees with the first normal mode of the embedded
cluster in limit free boundary condition. Further calculations indicate that
resonant mode at $\kappa _o<1$ is similar to that at $\kappa _o=0.1$ while
resonant mode at $\kappa _o>1$ is similar to that at $\kappa _o=2$. These
results indicate that resonance surely can occur in the embedded cluster
when the force constant $\kappa _o$ is sufficiently different from that of
the surrounding part and resonance frequencies shift a little from the
natural frequencies of embedded clusters with limit fixed or free boundary
conditions.

Another notable is resonance amplitude. It is obvious that resonance
amplitude $A_f$ increases linearly with the magnitude $f$ of the driving
force. It is found that resonance amplitude also depend on the size of the
embedded cluster. In Fig. 2(d) we plot $A_f$ at the fundamental frequency
versus $N_1$, and the curve shows $A_f$ increases linearly with $N_1$.
Moreover, $A_f$ decreases with the increase of $\kappa _o$ at $\kappa _o<1$.

We extend the 1D lattice to a 2D lattice. Fig. 2 (e) and (f) show the
vibration patterns at the fundamental frequency $\omega =0.026$ and the
second harmonic $\omega =0.041$, which correspond to the first and second
normal modes of the embedded cluster with limit fixed boundary condition
respectively. In the calculation the 2D lattice size is $100\times 100$ and
the embedded cluster at $\kappa _o=0.1$ is $50\times 50$ in size.

So it is concluded that resonance can occur in embedded clusters if their
force constants are sufficiently different from those of the surroundings
and driving frequencies match their natural frequencies. In the following we
illuminate the mechanism of Gao's finding in detail. First we discuss
whether there exist embedded clusters with quite different force constant in
materials. We come to a simple three-atom model to achieve the relationship
between the force constant $\kappa $ and the atomic separation $R$. Atoms $a$
and $c$ are fixed and atom $b$ is connected to them. The atomic separation
is $R$. Supposing that the interactions between atoms are the Lennard-Jones
potential $U=\sigma ^{12}R^{-12}-\sigma ^6R^{-6}$, then the force constant
of the atom $b$ is $\kappa \left( R\right) =-312\sigma ^{12}R^{-14}+84\sigma
^6R^{-8}$. When the separation $R$ increases to 1.4 times (a larger change
of lattice constant than 1.4 times has been observed in Gao's experiments%
\cite{7}), the force constant $\kappa _o$ decreases to about 0.06 times the
initial one. Therefore, a little increase of atomic separation can lead to
marked decrease of force constant and such embedded clusters can be
extensive in materials.

From equation (1) the natural frequency of the embedded cluster can be
estimated by $\omega \symbol{126}\frac 1{N_1}\sqrt{\frac{\kappa _o}m}\cos
\left( \frac{n\pi }{2N_1}\right) $ $\symbol{126}\frac 1{N_1a}\sqrt{\frac{%
a^2\kappa _o}m}\cos \left( \frac{n\pi }{2N_1}\right) $ $\symbol{126}$ $v/l$,
where $v$ verges on the sound velocity in the embedded cluster at around $%
10^3m/s$ and $l$ verges on the size of the embedded cluster at from $0.1\mu
m $ to $2\mu m$. Then the natural frequency is about $10^9$ $\symbol{126}$ $%
10^{10}$. It is obvious that in environment there exist external drives in
the frequency range, such as ubiquitous microwaves.

Next we study whether the resonance amplitude of the embedded cluster can be
large enough to be observed. Fig. 2(d) shows $A_f$ versus $N_1$ at $\kappa
_o=0.1$, $f=0.01$ and $T=0.01$. $A_f$ increases linearly with $N_1$ as $A_f$ 
$\symbol{126}$ $0.14N_1$. We make an approximate correspondence between
simulations and observations with the numerical result in fig. 2(d). For
example, we obtain $A_f$ $\symbol{126}$ $10^3$ at $N_1$ $\symbol{126}$ $10^4$%
. Then for such embedded cluster, the length is $1\mu m$ and the resonance
amplitude is about $10^2nm$, supposing that the atomic separation 1 in the
numerical simulations corresponds to 0.1 $nm$ in experiments. The value of
the resonance amplitude is comparable to that observed in Gao's finding. So
resonance in the embedded cluster is observable with an optical microscope.
Smaller localized clusters will be observed if the microscope has a larger
magnification.

In summary, we investigate mechanical resonance in the embedded cluster.
Resonance can take place when the structure of the embedded cluster is
different enough from the surrounding material. Moreover, we believe Gao's
finding is resonance of embedded clusters at the mesoscopic scale of
materials driven by ubiquitous microwaves. Embedded structures are common in
solids, so the phenomenon would have great significance in material physics.
Resonance in the embedded clusters may have influence on the thermodynamic
properties at the macroscale and fracture at the mesoscopic scale of
materials.

This work is supported by the National Natural Foundation of China, the
Major State Research Development 973 Project of Nonlinear Science in China.

\[
\text{Figure Captions} 
\]

Fig. 1

(a). The embedded clusters. (b). The 1D lattice model.

Fig. 2

(a). The average amplitude $A$ versus the frequency $\omega $ of the
external force. The values of the parameters are set as: $f=0.01$, $\kappa
_o=0.1$, $N=3000$ and $N_1=1000$. (b). The vibration pattern at $\omega
=0.80\times 10^{-3}$. $\kappa _o=0.1$. The $x$ axis is the site index $%
\left( i\right) $ of the atom; the $y$ axis is the atom displacement from
the equilibrium position. (c). The vibration pattern at $\omega =4.8\times
10^{-3}$. $\kappa _o=2$. (d) $A_f$ versus the size $N_1$ of the embedded
cluster. The values of the parameters are $f=0.01$, $\kappa _o=0.1$ and $%
N=3\times N_1$. (e) and (f). The two vibration patterns at the first and
second harmonic of the 2D model. The $x$ and $y$ axes show the site index $%
\left( i,j\right) $ of the atom; the $z$ axis shows the atom displacement
from the equilibrium position. The lattice size is $100\times 100$ and the
embedded cluster size is $50\times 50$. $\kappa _o=0.1$. $f=0.01$. $\omega
_1=0.026$ and $\omega _2=0.041$.

* zhaoh@xmu.edu.cn

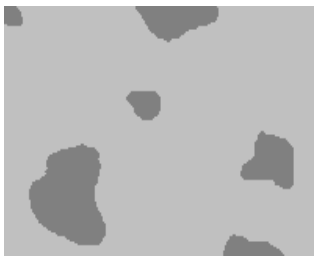
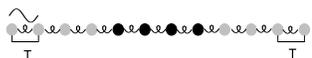
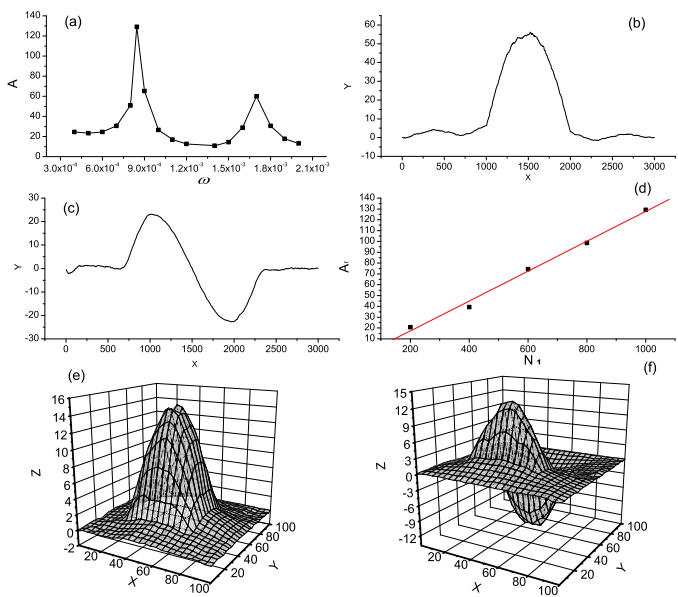
\end{document}